\documentclass[a4paper,11pt]{article}
\usepackage{jheppub}
\usepackage[utf8x]{inputenc}
\usepackage{graphicx}
\usepackage{float}
\usepackage[T1]{fontenc}
\usepackage{epsfig}
\usepackage{color}
\usepackage{tikz}
\usepackage{amsmath,mathtools}
\usepackage{subfloat}
\usepackage{amsfonts}
\usepackage{braket}
\usepackage{cleveref}
\usepackage{epstopdf}
\usepackage{caption}
\usepackage{subcaption}
\usepackage{enumitem}
\usepackage{etoolbox}
\usepackage{orcidlink}
\usepackage{comment}
\usepackage[titletoc,toc,title]{appendix}
\usepackage{hyperref}
\hypersetup{
	colorlinks=true,
	linkcolor=blue,
	filecolor=red,      
	urlcolor=blue,
	citecolor=red
}

\title{Holographic Entanglement Entropy in Janus deformed AdS$_3$ Geometries}

\author[]{Ankit Anand\orcidlink{0000-0002-8832-3212},\;}
\author[]{Himanshu Chourasiya\orcidlink{0000-0002-2774-7660},\;}
\author[]{Ankur Dey\orcidlink{0009-0001-1077-0442}\;}
\author[]{and Gautam Sengupta\orcidlink{0000-0002-1118-6926}}

\affiliation[]{
	Department of Physics,\\Indian Institute of Technology Kanpur,\\208016, India}
\emailAdd{anand@iitk.ac.in}
\emailAdd{chim@iitk.ac.in}
\emailAdd{ankurd21@iitk.ac.in}
\emailAdd{sengupta@iitk.ac.in}
\date{}

\abstract{

We investigate the time dependent entanglement entropy for boosted single intervals in interface conformal field theories (ICFT$_2$s) dual to Janus deformed AdS$_3$ geometries. For a Janus deformed Poincar\'e AdS$_3$ background, we obtain the entanglement entropy and a Janus induced correction using a replica technique for the equivalent dual field theory described on a conformally flat background on corresponding AdS$_2$ slices on the asymptotic boundary. The holographic entanglement entropy is then computed through certain embedding relations for the bulk Janus deformed AdS$_3$ geometry which exactly match with the field theory results. We further extend our analysis to investigate the entanglement entropy of corresponding intervals in ICFT$_2$s dual to bulk Janus deformed BTZ black hole and AdS$_3$ black string geometries obtaining consistent results from both the field theoretic and bulk computations.

}

\begin{document}

\maketitle


\section{Introduction}\label{sec_intro}

A class of asymptotically AdS spacetime described as Janus deformed geometries which are solutions of Einstein gravity with a non-trivial dilaton field profile were first introduced in the context of Type IIB supergravity on AdS$_3 \times S^3 \times \mathbb{M}_4$ \cite{Bak:2003jk, Freedman:2003ax, Bak:2007jm, Chiodaroli:2010ur}. In general, the bulk geometry is described by an AdS$_d$ sliced domain wall solution of an AdS$_{d+1}$ geometry mediated by a massless dilaton field with dependence on the slicing coordinate. Due to distinct asymptotic values of the dilaton on opposite halves of the boundary, a sudden transition is observed in the coupling constant of the corresponding marginal operator in the deformed dual field theory. This breaks the $SO(d,2)$ conformal symmetry down to $SO(d-1,2)$ preserved along the interface, thus realizing an interface conformal field theory (ICFT) as the holographic dual \cite{Clark:2004sb, Bak:2007jm, Gutperle:2016gfe}. An alternate effective lower dimensional description was subsequently developed in \cite{Bak:2020enw} to describe double sided Janus$_3$ black holes in the context of the island formalism and the black hole information loss paradox.

Early applications of Janus deformation within the AdS$_3$/CFT$_2$ correspondence focused on determining the boundary entropy contribution to the holographic entanglement entropy, which agreed with the ICFT computations using the $g$-theorem \cite{Azeyanagi:2007qj, Chiodaroli:2010ur, Estes:2014hka, Gutperle:2016gfe}. Subsequently this was extended to investigate bulk Janus deformed BTZ black hole dual to ICFTs at finite temperatures \cite{Bak:2011ga, Bak:2013uaa, Gutperle:2016gfe, Bak:2020enw} and to the time-dependent Janus deformed eternal BTZ black holes dual to two entangled thermal CFTs \cite{Bak:2007qw, Nakaguchi:2014eiu}. In recent works, the Janus deformation parameter has been analytically continued to imaginary values as a toy model of traversable wormholes and time-like Janus solutions dual to global quantum quenches of the corresponding dual ICFTs \cite{Kawamoto:2025oko, Suzuki:2025wlc}. Further studies have explored quantum information aspects such as complexity \cite{Stanford:2014jda, PhysRevD.93.086006, Emparan:2021hyr, 10.21468/SciPostPhys.6.3.034}, pseudo-entropy \cite{Nakata:2020luh}, and the emergence of islands \cite{Almheiri:2020cfm, Almheiri:2019hni, Almheiri:2019yqk,   Almheiri:2019qdq, Penington:2019npb, Penington:2019kki}, in the context of Janus deformed AdS geometries \cite{Bak:2020enw, Auzzi:2021nrj, Auzzi:2021ozb}. Check \cite{Goto:2020per, Afxonidis:2025jph, Baig:2024hfc, Gutperle:2022fma, Chen:2020efh, Chiodaroli:2016jod, Gutperle:2024yiz, Gutperle:2020gez, Gutperle:2017nwo, Karndumri:2025dqe, Karndumri:2024uxz, Karndumri:2021pva} for more interesting developments involving Janus deformed geometries.

Despite these advancements, existing analysis of entanglement entropy in Janus deformed AdS$_3$ geometries have been confined to purely spatial single intervals which are symmetric about the interface in the corresponding dual ICFTs. A natural question is whether these computations can be extended to more generic boosted single intervals which are asymmetric about the interface. Addressing this requires refinement of the usual bulk computation and the field theory techniques for Janus deformed geometries and the corresponding dual ICFTs which leads to interesting insights into the quantum information aspects of these models. 

In this article, we address this interesting question and obtain the holographic entanglement entropy for boosted single intervals asymmetric about the interface in ICFTs dual to Janus deformed AdS$_3$ geometries. In this context we first analyze the entanglement entropy of asymmetric single intervals in ICFTs dual to Janus deformed AdS$_3$ Poincar\'e geometry and the correction due to the marginal deformation. To this end the entanglement entropy is computed through a replica technique \cite{Calabrese:2009qy} utilizing a framework where the dual field theory is described on a conformally flat background arising from an AdS$_2$ foliation of the bulk geometry, as an alternative to the conventional ICFT description. Subsequently from the bulk perspective, we employ an embedding space formulation to compute the length of the RT/HRT surface \cite{Ryu:2006bv,Hubeny:2007xt} and obtain the holographic entanglement entropy with a Janus induced correction term. We observe an exact agreement between the field theory and the bulk computation results. Following this we apply our formulation to compute the modified holographic entanglement entropy for boosted single intervals in thermal ICFTs dual to bulk Janus deformed BTZ black hole and AdS$_3$ black string geometries respectively.

The remainder of the article is organized as follows. In \cref{sec_review}, we briefly discuss the essentials of Janus deformation and computations of the entanglement entropy for purely spatial interval symmetric about the interface in ICFTs dual to Janus deformed AdS$_3$ geometries. Subsequently, in \cref{sec_Poincare}, we describe the computation of the corrected entanglement entropy for generic boosted single intervals asymmetric about the interface in field theories dual to Janus deformed Poincar\'e AdS$_3$ geometry. Following this in \cref{sec_bs,sec_btz} we extend our analysis to Janus deformed BTZ black holes and AdS$_3$ black string geometries and their corresponding dual field theories. Finally, in \cref{sec_summary} we summarize our results and discuss their implications.


\section{Review of earlier literature}\label{sec_review}

In this section, we briefly review the construction of the three-dimensional Janus solution \cite{Bak:2007jm,Bak:2011ga} and its holographic dual described as an Interface CFT \cite{Bak:2020enw}. Subsequently we discuss the computation of the Janus induced correction to the entanglement entropy for a purely spatial single interval symmetric about the interface in the ICFT$_2$ dual to the bulk Janus deformed Poincar\'e AdS$_3$ geometry.


\subsection{Janus Deformed Solution}

\begin{figure}[t]
\centering
\includegraphics[scale=0.3]{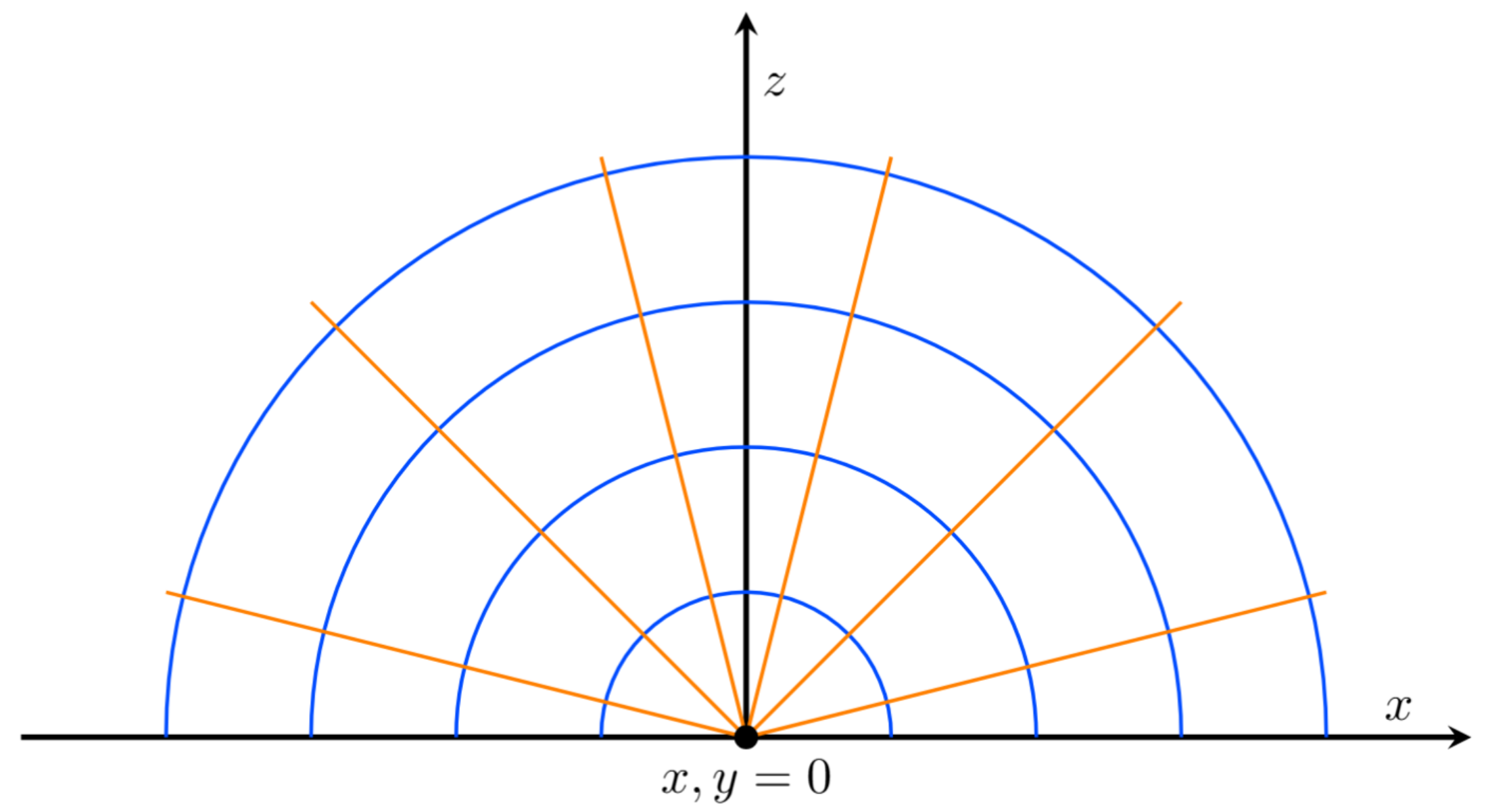}
\caption{This diagram representing the AdS$_2$ slicing of the AdS$_3$ Poincar\'e geometry. The orange lines represent the constant $\rho$ surfaces, where the induced metric on each slice is an AdS$_2$ geometry with a warping factor. The constant $y$ surfaces are represented by the blue curves. Figure modified from \cite{Auzzi:2021nrj}.} \label{fig_poincare}
\end{figure}

We now discuss the construction of a three-dimensional Janus solution \cite{Bak:2007jm,Bak:2011ga}, which can be embedded into a ten-dimensional type IIB supergravity on AdS$_3 \times S^3 \times \mathbb{M}_4$ described by the metric 
\begin{align}
    ds^2 = e^{\phi/2} (ds^2_3 + ds^2_{S_3}) + e^{-\phi/2} ds^2_{\mathbb{M}_4} ,
\end{align}
where $S_3$ is a 3-sphere, $\mathbb{M}_4$ is an internal manifold, and $\phi$ is a non-trivial dilaton profile along the AdS$_3$ directions introduced by the deformation. For non-supersymmetric Janus solutions the dilaton and the metric $g_{ab}$ of AdS$_3$ geometry are independent of the coordinates of $S_3$ and $\mathbb{M}_4$. As a result, through a dimensional reduction the relevant dynamics of $\phi$ and $g_{ab}$ may be described by the action
\begin{align}
    I = \frac{1}{16\pi G_3} \int d^3x \,\sqrt{-g}\,\left( R_3 - g^{ab}\partial_a\phi\,\partial_b\phi + \frac{2}{\ell^2} \right) ,
\end{align}
which couples the three-dimensional Einstein gravity with a negative cosmological constant to a scalar field. The variation of the above action with respect to the metric leads to the Einstein field equation
\begin{align}\label{eom1}
    R_{ab} + 2 g_{ab} = \partial_{a}\phi \, \partial_{b}\phi ,
\end{align}
while the variation with respect to $\phi$ results in the equation of motion of the dilaton field as
\begin{align}\label{eom2}
    \nabla_{a}\nabla^{a}\phi \equiv \frac{1}{\sqrt{-g}} \partial_{a}\!\left( \sqrt{-g}\, g^{ab}\, \partial_{b}\phi \right) = 0  .
\end{align}
The above equations may be solved by adopting a metric ansatz for a Janus deformed AdS$_3$ geometry as
\begin{align}\label{Janus Metric}
    ds^2_3 =  f(\rho)\, ds^2_{\text{AdS}_2} + d\rho^2  , \qquad \phi = \phi (\rho) ,
\end{align}
where $ds^2_{\text{AdS}_2}$ denotes the AdS$_2$ line element and $\rho \in (-\infty, +\infty)$ is a hyperbolic angular coordinate such that the asymptotic boundary lies at $\rho \to \pm \infty$.\footnote{Note that this hyperbolic angle $\rho$ is different in nature from the hyperbolic angle for an undeformed AdS$_3$ geometry, and are related via a finite multiplicative factor dependent on the deformation parameter $\gamma$ for large hyperbolic angles.} The Janus deformed AdS$_3$ geometry can thus be interpreted as an AdS$_2$ foliation of AdS$_3$, where each slice is warped by a non-trivial function of $\rho$ (and as we shall see, on the deformation parameter $\gamma$ as well). In case of a Janus deformed Poincar\'e AdS$_3$ geometry (depicted in \cref{fig_poincare}), the above metric becomes
\begin{align}\label{Poincare_metric}
    ds_3^2 =  f(\rho) \frac{-dt^2+dy^2}{y^2}+d \rho^2  ,
\end{align}
where $t$ is the Lorentzian time coordinate and $y$ is a radial coordinate extending into the bulk. Solving the equations of motion in \cref{eom1,eom2} using the ansatz in \cref{Janus Metric} yields the solution 
\begin{align}\label{f(rho)_def}
    f(\rho) & = \frac{1}{2}\left(1 + \sqrt{1 - 2\gamma^2}\,\cosh2 \rho \right), \notag \\ \phi(\rho) & = \phi_0 + \frac{1}{\sqrt{2}} \log \left(\frac{1 + \sqrt{1 - 2\gamma^2} + \sqrt{2}\gamma \tanh \rho}{1 + \sqrt{1 - 2\gamma^2} - \sqrt{2}\gamma \tanh \rho}\right)  ,
\end{align}
where the integration constant $\phi_0$ is the value of the dilaton field at $\rho=0$, while $\gamma \in [0, \frac{1}{\sqrt{2}}]$ is the Janus deformation parameter. The undeformed AdS$_3$ geometry corresponds to $\gamma = 0$, where the dilaton field is now a global constant $\phi_0$.

The holographic dual of the Janus deformed AdS$_3$ geometry is a CFT$_2$ deformed by a marginal operator $\mathcal{O}(w,\bar{w})$. In the conformal perturbative limit, the action of the deformed theory is described as
\begin{align}
    I_{\gamma} = I_0 + \gamma \int  d^2w ~J_{\pm} ~\mathcal{O}(w,\bar{w}) + O(\gamma^2),
\end{align}
where $w,\bar{w}$ represent the complex coordinate and its conjugate in the undeformed field theory on the asymptotic boundary. The couplings $J_{\pm}$ are related to the asymptotic values of the dilaton field $\phi$ as $J_{\pm}=\lim_{\rho \to \pm \infty} \phi(\rho)$. Given that there are two different asymptotic values of $\phi$, the resultant theory is an interface CFT$_2$ where different coupling constants are observed on either side of the interface. The marginal nature of the deformation preserves the central charge, and the resulting theory can be schematically represented as \cite{Bak:2020enw}
\begin{align}
	\text{ICFT}= \text{CFT}_{-} \times \text{QM} \times \text{CFT}_{+},
\end{align}
where “QM” denotes the quantum mechanical system of the interface degrees of freedom. The presence of the interface breaks the $SO(d,2)$ conformal symmetry of the undeformed CFT$_2$. Consequently the ICFT$_2$ preserves a $SO(d-1,2)$ conformal symmetry along the interface.


\subsection{Correction to entanglement entropy}

We now discuss the computations of the Janus induced correction to the entanglement entropy. Considering a CFT$_2$ defined on a spatial manifold partitioned into two complementary regions $A$ and $B$, the complete Hilbert space factorizes as
\begin{equation*}
	\mathcal{H} = \mathcal{H}_A \otimes \mathcal{H}_B  ,
\end{equation*}
where $\mathcal{H}_{A,B}$ correspond to the subspaces associated with $A$ and $B$, respectively. The reduced density matrix for region $A$ is obtained by tracing out the degrees of freedom in $B$
\begin{equation*}
	\rho_A = \mathrm{Tr}_{\mathcal{H}_B} \, \rho_{AB}  ,
\end{equation*}
where $\rho_{AB}$ denotes the density matrix of the full system. The entanglement entropy of region $A$ is then defined as
\begin{equation*}
	S_A = - \mathrm{Tr}_{\mathcal{H}_A} (\rho_A \log \rho_A)  .
\end{equation*}
While the computation of entanglement entropy is relatively straightforward for finite quantum systems, it becomes computationally intractable in quantum field theories (QFTs), where the reduced density matrix is infinite-dimensional. The authors in \cite{Calabrese:2004eu, Calabrese:2009ez, Calabrese:2009qy} developed a replica technique to compute the entanglement entropy in CFT$_2$ using the R\'enyi generalization of the entanglement entropy as 
\begin{equation} \label{Renyi-EE-def}
	S_A =  \lim _{n \to 1} S_{n}  = \lim _{n \to 1} \frac{1}{1-n} \log \text{Tr} (\rho_A)^n  .
\end{equation}

 The field theory computation of the Janus induced correction to the entanglement entropy of a subsystem $A$ assumed on the asymptotic boundary is described in details in \cite{Azeyanagi:2007qj}. Assuming the internal manifold $\mathbb{M}_4$ to be a 4-torus $\mathbb{T}_4$, the authors state that the correction to the entanglement entropy for a purely spatial interval symmetric about the interface in an ICFT dual to a Janus deformed AdS geometry characterizes the degrees of freedom localized at the interface, and may be described by a universal term 
\begin{align}
    \Delta S_A \propto \log \frac{\sqrt{\frac{R_{+}}{R_{-}}+\frac{R_{-}}{R_{+}}}}{\sqrt{2}}  .
\end{align}
Here $R_{\pm}$ describes the compactified radius of the 4-torus (with different radii on either side of the interface) related to the asymptotic values $\phi_{as}$ of the dilaton field. Given that $\phi_{as}$ is in turn related to the deformation parameter $\gamma$, the ratios of the two radii may be described as
\begin{align}
    \frac{R_{+}}{R_{-}} = \left( \frac{1+\sqrt{2}\gamma}{1-\sqrt{2}\gamma} \right)^{\frac{1}{2 \sqrt{2}}}  .
\end{align}
Finally, the correction to the entanglement entropy is then perturbatively obtained as
\begin{align}\label{cft_corr}
    \Delta S_A = \frac{c}{6} \gamma^2 + \mathcal{O}(\gamma^4) ,
\end{align}
where $c$ describes the central charge of the field theory.

\vspace{0.5cm}
In the AdS$_3$/CFT$_2$ framework, the holographic entanglement entropy is computed using the Ryu-Takayanagi (RT) prescription \cite{Ryu:2006bv, Ryu:2006ef}. For a spatial interval $A$ considered on the asymptotic boundary of a bulk AdS$_3$ geometry, the holographic entanglement entropy is obtained in terms of the area $\mathcal{L}$ of the minimal surface homologous to $A$ as 
\begin{align}\label{eq_rt}
    S_A = \frac{\mathcal{L}}{4 G_3} .
\end{align}

The correction to the holographic entanglement entropy induced by Janus deformation in Poincar\'e AdS$_3$ geometries is derived in \cite{Azeyanagi:2007qj}. Once again, considering that $A$ is symmetric about the interface and has endpoints at $(t,y_1,\pm \rho_c)$, the minimal surface can be parametrized in terms of only the hyperbolic coordinate $\rho$, which varies from $-\rho_c$ to $+\rho_c$. In the vicinity of the asymptotic boundary, the metric in \cref{Poincare_metric} becomes 
\begin{align}\label{Metric_at_rhoc}
    ds^2 = d \rho^2 + \sqrt{1-2\gamma^2}~\frac{e^{2\rho}}{4} \frac{-dt^2+dy^2}{y^2}  .
\end{align}

Evaluation of the holographic entanglement entropy \cite{Azeyanagi:2007qj, Chiodaroli:2010ur, Bak:2011ga} requires the computation of the regularized geodesic length near the asymptotic boundary. Given that the Janus deformed geometry is asymptotically AdS$_3$, this may be achieved by equivalently expressing the asymptotic geometry as 
\begin{align}
    ds^2 \simeq \frac{1}{z^2}\left(dz^2 + dx^2 - dt^2\right)  ,
\end{align}
where $z = \epsilon$ is the UV regularization of the CFT at the asymptotic boundary. This equivalence further allows us to determine the hyperbolic angles $\rho_c$ in terms of $\epsilon$ as 
\begin{align}
   \left(1-2\gamma^2\right)^{-\tfrac{1}{4}} \frac{e^{\rho_c}}{2} = \frac{y_1}{\epsilon} \implies \rho_c = \log{\left[\frac{2y_1}{\epsilon}\,\left(1-2\gamma^2\right)^{-\tfrac{1}{4}}\right]}  ,
\end{align}
Using the above relation, the correction to the geodesic length due to the Janus deformation may be evaluated as
\begin{align}\label{eq:DeltaL}
\Delta \mathcal{L} \equiv \rho_c^{(\gamma)} - \rho_c^{(0)} =  -\frac{1}{2} \log (1 - 2\gamma^2) .
\end{align}
Using the RT formula, the correction to the holographic entanglement entropy due to Janus deformation may then be determined as
\begin{align}\label{eq:DeltaSA}
\Delta S_A = \frac{\Delta \mathcal{L}}{4G_3} = \frac{1}{4G_3}\gamma^2  +\mathcal{O}(\gamma^4).
\end{align}
which matches with \cref{cft_corr} on applying the Brown Henneaux relation $c = \frac{3}{2 G_3}$ \cite{Brown:1986nw}.


\section{Janus deformed Poincar\'e AdS$_3$ geometry}\label{sec_Poincare}

Having discussed the Janus deformed AdS$_3$ solutions and computation of the entanglement entropy for single spatial intervals symmetric about the interface, we now consider boosted single intervals asymmetric about the interface in the field theory dual to the Janus deformed Poincar\'e AdS$_3$ geometry described by \cref{Poincare_metric}. In this case, the metric on each foliation (induced on a constant $\rho$ slice) is a Poincar\'e AdS$_2$ geometry warped by $f(\rho)$. As described previously, the asymptotic boundary lies at $\rho \to \pm \infty$.

We begin with the field theoretic computation of the entanglement entropy. The key step is the conformal mapping of the effective boundary metric induced by the bulk geometry to a conformally flat form. The entanglement entropy is then evaluated using the replica technique, where the Janus induced correction to the entanglement entropy arises from the Weyl conformal factor. From the bulk perspective, we utilize certain embedding relations to determine the length of the RT/HRT surface and obtain the holographic entanglement entropy.


\subsection{Entanglement Entropy} \label{ssec_pcft}

\begin{figure}[t]
\centering
\includegraphics[scale=0.5]{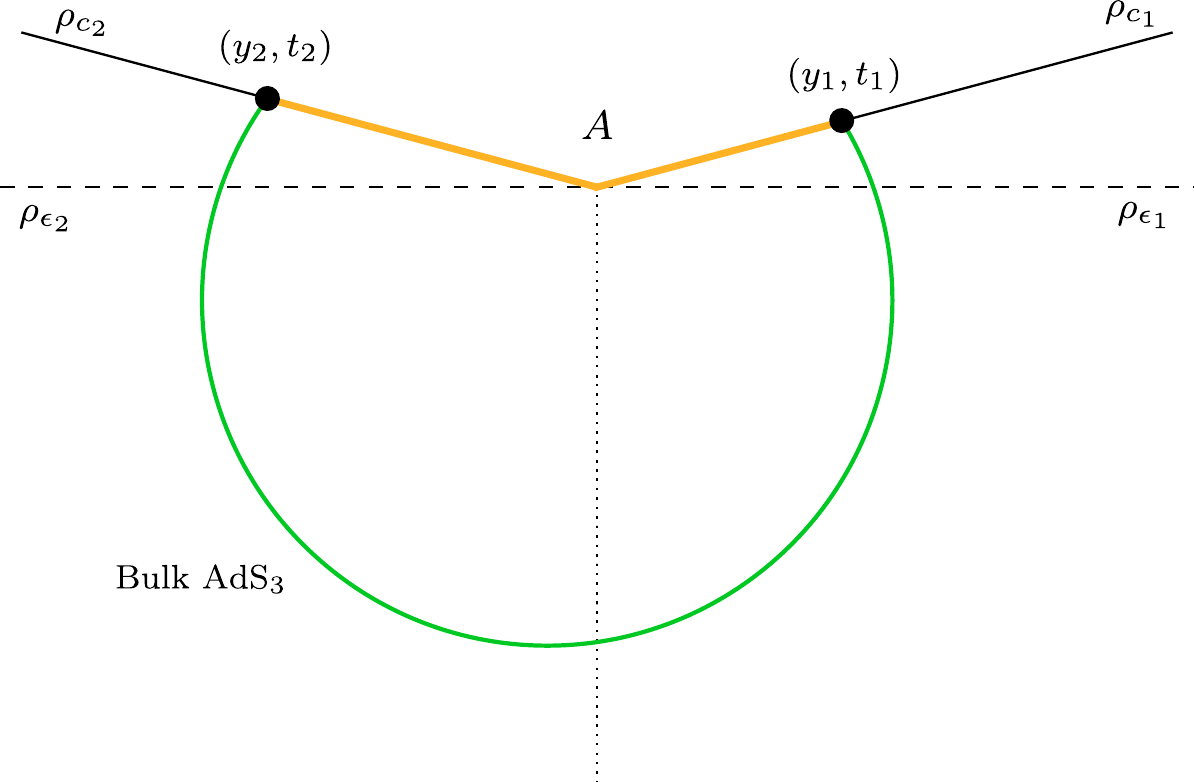}
\caption{This diagram illustrates a boosted single interval $A$ (depicted in orange) asymmetric about the interface in the field theory dual to the bulk Janus deformed Poincar\'e AdS$_3$ geometry. The solid black lines represent the asymptotic boundary of the Janus deformed geometry, while the black dashed lines denote the asymptotic boundary for an undeformed geometry. Additionally, the green curve represents the minimal geodesic (the RT/HRT surface) between the endpoints of the subsystem $A$.} \label{fig_subsystem}
\end{figure}

In this subsection, we consider a boosted single interval $A=[(y_1,t_1),(y_2,t_2)]$ on the asymptotic boundary of a Janus deformed Poincar\'e AdS$_3$ geometry (as illustrated in \cref{fig_subsystem}), and compute the entanglement entropy between $A$ and its compliment using the replica trick \cite{Calabrese:2009qy}. The metric induced at the asymptotic boundary $\rho=\rho_c$ of the bulk geometry is given
\begin{align}
	ds^2 = f(\rho_c)\left(\frac{-dt^2 + dy^2}{y^2}\right)  ,
\end{align}
which may further be expressed as a conformally flat form as
\begin{align}\label{eq_cf}
ds^2 &=	\sqrt{1-2 \gamma ^2} \left(\frac{e^{\rho_c}}{2 y}\right)^2 \left(-d t^2+d y^2\right) = \Omega^2(y) \, ds_{\text{flat}}^2  ,
\end{align}
where $\Omega(y)= \frac{(1-2 \gamma ^2)^\frac{1}{4} }{2 y}e^{\rho_c}$ is the conformal factor. Utilizing the replica trick, the entanglement entropy for a subsystem $A$ may be computed as
\begin{align}\label{Def_S(A)_twist}
	S(A)= \lim_{n\to 1} \frac{1}{1-n} \log \Omega(y)^{-\Delta_n} \langle \phi_n (y_1,t_1) \phi_n (y_2,t_2) \rangle ,
\end{align}
where $\phi_n (y_1)$ and $\phi_n (y_2)$ are the twist field operators located at the end points of the subsystem $A$, while $\Delta_n=\frac{c}{12}\left(n-\frac{1}{n}\right)$ is the conformal dimension of the twist operators. Now using the known form of the two-point function\footnote{Here we utilize the fact that the $y$ coordinate is a radial coordinate and the subsystem $A$ is defined about the interface, making the length of the subsystem ($y_1+y_2$).} and subsequently taking the replica limit, we can obtain 
\begin{align}\label{PEE-Foliation}
	 S(A)&=\frac{c}{6}  \log \left[\frac{e^{\rho_{c_{1}}} e^{\rho_{c_{2}}}\left((y_2+y_1)^2+(t_2-t_1)^2\right)}{4 y_1 y_2}\right]+ \frac{c}{12}  \log \left(1-2 \gamma ^2 \right) ,\notag\\
     &=S_{\gamma}(A)+ \frac{c}{12}  \log \left(1-2 \gamma ^2 \right)  .
\end{align}
where $S_{\gamma}(A)$ represents the Janus deformed entanglement entropy expressed in terms of the deformed hyperbolic coordinates $\rho_{c_1}$ and $\rho_{c_2}$. Putting $\gamma=0$ leads to the undeformed scenario, and using \cref{PEE-Foliation} we have
\begin{align}
     S(A) = \frac{c}{6}  \log \left[\frac{e^{\rho_{\epsilon_{1}}} e^{\rho_{\epsilon_{2}}}\left((y_2+y_1)^2+(t_2-t_1)^2\right)}{4 y_1 y_2}\right]  ,
\end{align}
where $\rho_{\epsilon_{1}}$ and $\rho_{\epsilon_{2}}$ are the hyperbolic coordinates in the undeformed AdS$_3$ and are related to the UV cut-off as
\begin{align}\label{IRtoUV}
	\rho_{\epsilon_{1}} = \log \frac{2 y_1}{\epsilon} , \qquad\rho_{\epsilon_{2}} = \log \frac{2 y_2}{\epsilon} .
\end{align}
The entanglement entropy in the Janus deformed Poincar\'e AdS$_3$ geometry may then be expressed in terms of an undeformed entanglement entropy and a correction term
\begin{align}\label{AdS Final length}
    S_{\gamma}(A)  = \frac{c}{6} \log\!\left[\frac{e^{\rho_{\epsilon_1}+\rho_{\epsilon_2}}\left(-(t_1-t_2)^2+(y_1+y_2)^2\right)}{4\,y_1\,y_2}\right] - \frac{c}{12} \log {(1-2\gamma^2)} .
\end{align}

From the above expression, the results provided in \cite{Azeyanagi:2007qj,Bak:2011ga} may be reproduced by considering $y_1=y_2=L/2$ and $t_1=t_2$, which gives us
\begin{align}
   S_{\gamma}(A)= \frac{c}{3}  \log \left(\frac{L}{\epsilon}\right) - \frac{c}{12}  \log \left(1-2 \gamma ^2 \right) .
\end{align}


\subsection{Holographic Entanglement Entropy}\label{ssec_pbulk}

To compute the holographic entanglement entropy, we begin with the Janus deformed Poincar\'e AdS$_3$ metric described in \cref{Poincare_metric}, which near the asymptotic boundary takes the form
\begin{align}
      ds^2 =  f(\rho)\, \left(\frac{-dt^2 + dy^2}{y^2}\right) + d\rho ^2, \qquad f(\rho) = \sqrt{1-2\gamma^2} ~ \frac{e^{2 \rho}}{4}  .
\end{align}
Given that Janus deformed geometries are asymptotically AdS in nature, the above metric may be systematically constructed using the embedding space formalism. In this framework, this spacetime may be embedded as a codimension-one hyperboloid in an $\mathbb{R}^{(2,2)}$ flat geometry using the embedding relations given as
\begin{align} \label{AdS Embedding}
    T_{1}(t,\,y,\,\rho)= \frac{1-t^2+y^2}{2y}\sqrt{f(\rho)} , \qquad & \qquad T_{2}(t,\,y,\,\rho)= \frac{t}{y}\sqrt{f(\rho)}, \notag\\
    X_{1}(t,\,y,\,\rho)=\frac{1+t^2-y^2}{2y}\sqrt{f(\rho)}  , \qquad & \qquad X_{2}(t,\,y,\,\rho)=\sqrt{f(\rho)-1}  ,
\end{align}
with an additional constraint describing a codimension-1 hyperboloid in $\mathbb{R}^{(2,2)}$ as
\begin{align}\label{eq_constraint}
    -T_1^2 - T_2^2 + X_1^2 + X_2^2 = -1 .
\end{align}
The AdS metric can be recovered from the embedding relations in \cref{AdS Embedding} by taking the large-$\rho$ limit within the embedding space formalism, wherein the ambient flat spacetime is equipped with the metric
\begin{align}\label{AdS Embedd Metric}
    ds^{2} = -dT_{1}^{2} - dT_{2}^{2} + dX_{1}^{2} + dX_{2}^{2}  .
\end{align}
An important advantage of this formulation lies in its ability to evaluate geometric quantities, such as the lengths of extremal surfaces, without the need to explicitly solve the geodesic equations. The geodesic distance between two bulk points is then encoded in the scalar product of their position vectors in the embedding coordinates. Consequently, the length $\mathcal{L}$ of a geodesic connecting two arbitrary points with embedding coordinates $(T_{1},\,T_{2},\,X_{1},\,X_{2})$ and $(T_{1}',\,T_{2}',\,X_{1}',\,X_{2}')$ is expressed as
\begin{align}\label{AdS Length}
    \mathcal{L} = \cosh^{-1}\!\Big( T_{1}T_{1}' + T_{2}T_{2}' - X_{1}X_{1}' - X_{2}X_{2}' \Big) .
\end{align}
Using \cref{AdS Length} and the embedding relations in \cref{AdS Embedding}, the geodesics length between two arbitrary bulk points at $(t_1,\,y_1,\,\rho_1)$ and $(t_2,\,y_2,\,\rho_2)$ may be obtained as 
\begin{align}\label{length_Poincare}
    \mathcal{L} = \cosh^{-1}\!\left[\frac{e^{\rho_1+\rho_2}\left(-(t_1-t_2)^2+(y_1+y_2)^2\right) \sqrt{1-2\gamma^2}}{8\,y_1\,y_2}\right]   .
\end{align}

Considering that the endpoints of the subsystem on the asymptotic boundary to be at $(t_1,\,y_1,\,\rho_{c_1})$ and $(t_2,\,y_2,\,\rho_{c_2})$, using \cref{length_Poincare} the holographic entanglement entropy may be determined as 
 \begin{align}\label{final_length_Ads3}
 S_A & = \frac{1}{4G_3}\log\!\left[\frac{e^{\rho_{c_1}+\rho_{c_2}}\left(-(t_1-t_2)^2+(y_1+y_2)^2\right) \sqrt{1-2\gamma^2}}{4\,y_1\,y_2}\right] \notag \\ & = S_\gamma(A) + \frac{1}{8G_3}\log{(1-2\gamma^2)}  ,
\end{align}
where we utilize the identity $\cosh^{-1}{x} \sim \log{2x}$ for large $x$.

Following the discussion presented in \cref{ssec_pcft}, the holographic entanglement entropy in the Janus deformed Poincaré AdS$_3$ geometry in terms of the undeformed entanglement entropy and a correction term may be expressed as 
\begin{align}
    S_{\gamma}(A)  = \frac{1}{4G_3} \log\!\left[\frac{e^{\rho_{\epsilon_1}+\rho_{\epsilon_2}}\left(-(t_1-t_2)^2+(y_1+y_2)^2\right)}{4\,y_1\,y_2}\right] - \frac{1}{8G_3} \log {(1-2\gamma^2)}.
\end{align}

Additionally, the results for the Janus deformed holographic entanglement entropy for the symmetric configuration described in \cite{Bak:2011ga, Azeyanagi:2007qj} can be reproduced from \cref{AdS Final length} by using the relations in \cref{IRtoUV}, setting $(t_2=t_1,y_2=y_1)$, and finally using the Brown Henneaux relation $c = \frac{3}{2 G_3}$ \cite{Brown:1986nw}.

\vspace{1cm}

It is important to discuss some subtle points about the field theory and bulk computations for the entanglement entropy described above. Firstly, the Janus deformed geometry described in \cref{Janus Metric} is only asymptotically AdS for large values of the hyperbolic angle $\rho$ (i.e. in the vicinity of the asymptotic boundary, as elaborated in \cref{ssec_pcft,ssec_pbulk}). This allows us to utilize the AdS/CFT correspondence in an otherwise near-AdS spacetime, simplifying the computations of the entanglement entropy in such deformed scenarios.

Secondly, we utilize two different angular coordinates, namely $\rho_c$ and $\rho_{\epsilon}$, which describe the location of the asymptotic boundary in the Janus deformed and the undeformed spacetime, respectively. Consequently, any quantity dependent on $\rho_c$ ($S_{\gamma}(A)$ for instance) is defined on the Janus deformed spacetime, while those described in terms of $\rho_{\epsilon}$ (for example $S(A)$) are defined on an undeformed AdS spacetime. $\rho_c$ is intrinsically dependent on the deformation parameter $\gamma$ and approaches $\rho_{\epsilon}$ as $\gamma \to 0$.

Finally, as evident from \cref{eq_cf}, the Weyl conformal factor for Janus deformed theories dependent explicitly on the deformation parameter $\gamma$. As a result, in the field theoretic framework described above, the correction to the entanglement entropy originates from this modified conformal factors.


\section{Janus deformed planar BTZ black hole geometry }\label{sec_btz}

\begin{figure}[t]
\centering
\includegraphics[scale=0.33]{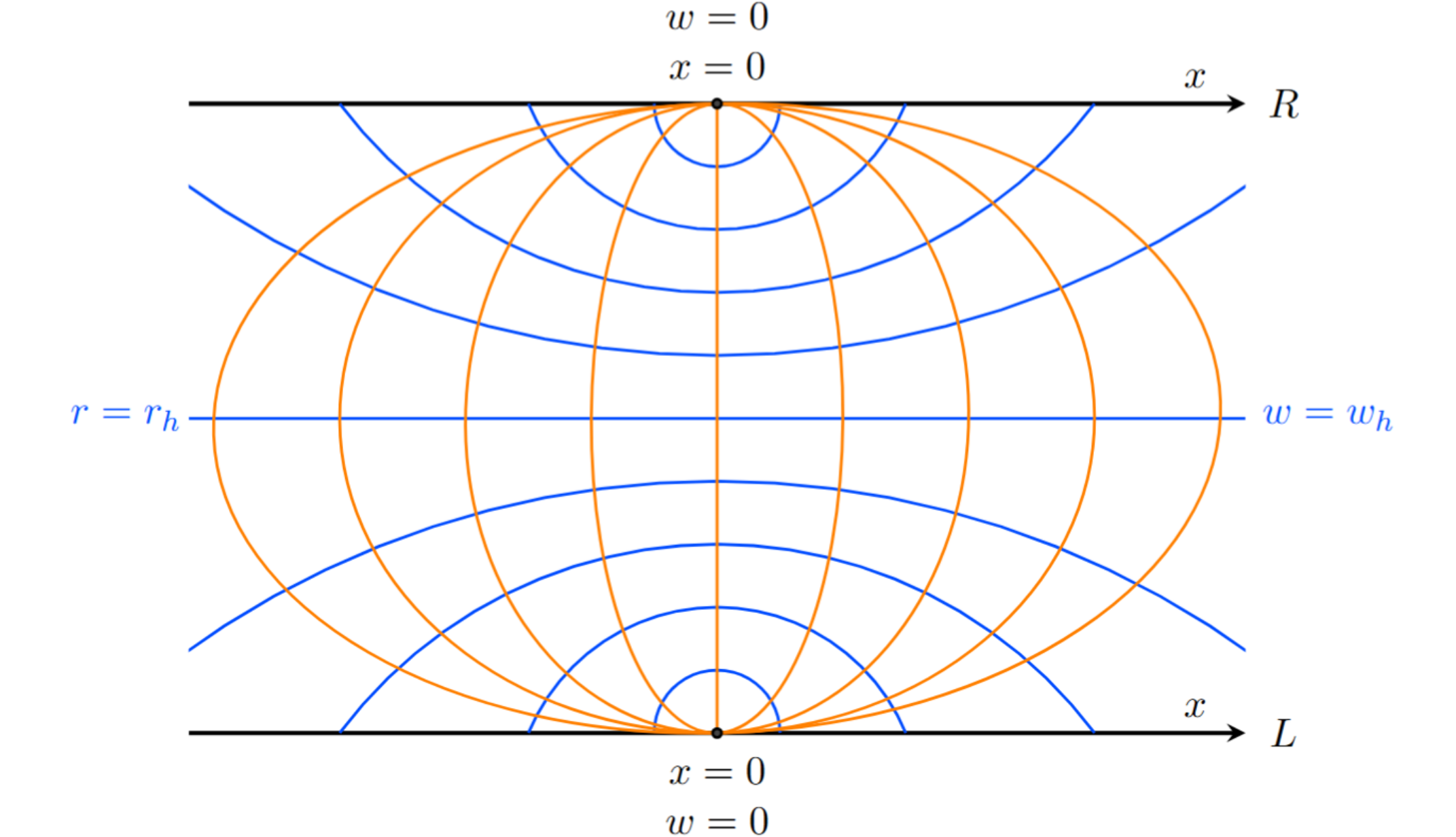}
\caption{ This diagram representing the AdS$_2$ slicing of the BTZ black hole geometry. The constant $\rho$ curves are depicted in orange, while the blue curves represent constant $w$ curves. Figure modified from \cite{Auzzi:2021nrj}.} \label{fig_btz}
\end{figure}

We now extend our analysis to Janus deformed BTZ black holes, also termed as Janus black holes. The metric for an undeformed planar BTZ black hole is given as \cite{Banados:1992wn} 
\begin{align}\label{eq_BTZ}
    ds^2 = -(r^2-r_h^2)dt^2 + \frac{dr^2}{r^2-r_h^2}+r^2dx^2 ,
\end{align}
where $r$ is the bulk radius coordinate, $t$ denotes the usual time coordinate, while $x$ is the spatial coordinate ranging from $- \infty$ to $+ \infty$. The horizon of the black hole is located at $r=r_h$, while the asymptotic boundary is located at $r \to \infty$. Utilizing the following transformations
\begin{align}\label{btz_trans}
    r = r_h \, \cosh{\rho} \;\sqrt{\frac{\omega_h^2}{\omega^2}-\tanh^2{\rho}} , \qquad \sinh{r_hx} = \frac{\tanh{\rho}}{\sqrt{\frac{\omega_h^2}{\omega^2}-\tanh^2{\rho}}}  ,
\end{align}
we can then express the metric in \cref{eq_BTZ} in terms of the AdS$_2$ foliation coordinates $(t, w, \rho)$ as
\begin{align}\label{eq_BTZfoliation}
    ds^2 = d\rho^2 + \cosh^2{\rho}  \left[-\frac{\;r_h^2(\omega_h^2-\omega^2)}{\omega^2}dt^2+\frac{\omega_h^2}{\omega^2(\omega_h^2-\omega^2)}d\omega^2\right] ,
\end{align}
where $w$ is the radial coordinate increasing into the bulk and $\rho$ is once again the hyperbolic angular coordinate defined earlier (see \cref{fig_btz}). The black hole horizon in this coordinate system is now located at $\omega = \omega_h$. Recall that the asymptotic boundary is located at $\rho = - \infty \cup \infty$. 

The Janus black hole may now be expressed in terms of the foliation coordinates as \cite{Bak:2011ga, Bak:2020enw}
\begin{align}\label{BTZ_metric}
    ds^2 = d\rho^2 + f(\rho)\; \left[-\frac{\;r_h^2(\omega_h^2-\omega^2)}{w^2}dt^2+\frac{\omega_h^2}{\omega^2(\omega_h^2-\omega^2)}d\omega^2\right]  ,
\end{align}
where the wrapping factor $f(\rho)$ is defined earlier in \cref{f(rho)_def}. The dual field theory to the Janus black hole solution is an ICFT at a finite temperature.

 For $\gamma=0$, the solution reduces to the undeformed BTZ black hole~\cite{Banados:1992gq}, where the horizon lies at $r=r_{h}$ and $r=0$ corresponds to an orbifold/conical singularity. Such a solution is invariant under translations in both $t$ and $x$, allowing the identification $x\sim x+2\pi$. For $\gamma\neq 0$ however, translational invariance along $x$ is broken due the dilaton field $\phi$ having two distinct asymptotic values while time-translation symmetry persists, reflecting the interface nature of the dual CFT.

 We first discuss the field theory computations of the entanglement entropy of a boosted single interval asymmetric about the interface on the asymptotic boundary of the Janus black hole geometry following a methodology similar to that described in \cref{ssec_pcft}. Subsequently, we describe the bulk computations using the embedding relations formalism.


\subsection{Entanglement Entropy} \label{ssec_btzcft}
In this subsection, we describe the computation of the entanglement entropy between a subsystem $A=[(\omega_{1},t_1),(\omega_{2},t_2)]$ and its complement in the Janus deformed BTZ black hole, utilizing the replica technique. Utilizing \cref{BTZ_metric} the metric at the asymptotic boundary in this scenario is given as
\begin{align}
	ds^2 = f(\rho_c )\left[-\frac{r_{h}^2 \left(\omega_{h}^2-\omega ^2\right)}{\omega ^2} dt^2+\frac{\omega_{h}^2}{ \omega^2 \left(\omega_h ^2-\omega^2\right)} d\omega ^2\right] .
\end{align}
The entanglement entropy between a subsystem $A$ and its complement may be computed from the two-point twist field correlator using as
\begin{align}\label{BTZ_correlators}
	S(A)= \lim_{n\to 1} \frac{1}{1-n} \log  \langle \phi_n (\omega_{1}) \phi_n (\omega_{2}) \rangle  .
\end{align}
 Since the field theory is not conformally flat, the computation of the two-point twist field correlator is not straightforward. To proceed, one first maps the theory to a flat background through a series of conformal transformations,
\begin{align}\label{btz_tr1}
	t = \frac{1}{2 r_{h}}  \log \frac{\bar{\eta}}{\eta }~~, \quad\quad  ~\omega = \omega_{h} \left(\frac{\eta~\bar{\eta}-\omega^2_{h}}{\eta~\bar{\eta}+\omega^2_{h}} \right).
\end{align}
In these coordinates, the metric becomes conformally flat,
\begin{align}
	ds^2=\frac{4 \omega_{h}^2 f(\rho ) }{\left(\omega_{h}^2-\eta  ~\bar{\eta}\right)^2}~d\eta  ~d\bar{\eta}  .
\end{align}
It is further possible to map this metric to the upper half plane through the transformations
\begin{align}\label{btz_tr2}
	\eta = \frac{\omega_{h}}{\nu -\frac{i}{2}}-i \omega_{h}, \qquad \bar{\eta} = \frac{\omega_{h}}{\bar{\nu}+\frac{i}{2}}+i \omega_{h} ,
\end{align}
where the metric transforms to
\begin{align}
	ds^2= \frac{4 f(\rho ) }{(\nu -\bar{\nu})^2}~ d\nu  ~d\bar{\nu} .
\end{align}
The corresponding conformal factor is given as 
\begin{align}
	\Omega(\nu, \bar{\nu})=\frac{2 \sqrt{f(\rho )} }{(\nu -\bar{\nu})}  .
\end{align}
Using this mapping, the twist field correlator \cref{BTZ_correlators} may be written as flat plane twist field correlator with an appropriate conformal factor as 
\begin{align}
	S(A)= \lim_{n\to 1} \frac{1}{1-n} \log \Omega(\nu,\bar{\nu})^{-\Delta_n} \langle \phi_n (\nu_1,\bar{\nu}_1) \phi_n (\nu_2,\bar{\nu}_2) \rangle_{\text{Flat}}.
\end{align}
Using the standard form of the two point function and the conformal factor, the entanglement entropy is given as
\begin{align}
	S(A)=\lim_{n\to 1}\frac{1}{1-n}~\log \left[\frac{(\Omega_1 \Omega_2)^{-\Delta_n }}{(\nu_1-\bar{\nu}_{2})^{2 \Delta_n }}\right]  .
\end{align}
Substituting the conformal factor and subsequently taking the replica limit, the above expression may be written as
\begin{align}
	S(A)=\frac{c}{6} \log \left[\frac{(\nu_1-\bar{\nu}_{2})^2 e^{\rho_{c_1}+\rho_{c_2}}}{(\nu_1-\bar{\nu}_{1}) (\nu_2-\bar{\nu}_{2})}\right]+\frac{c}{12}  \log \left(1-2 \gamma ^2\right)  .
\end{align}
Finally, transforming back the above result to the foliation coordinates using \cref{btz_tr1,btz_tr2}, we have
\begin{align}\label{ee_btz}
	S(A)=\frac{c}{6}  \log \left[\frac{ \left(\omega_{1} \omega_{2}+\omega_{h}^2-\sqrt{\left(\omega_{1}^2-\omega_{h}^2\right) \left(\omega_{2}^2-\omega_{h}^2\right)} \cosh r_{h}(t_2-t_1)\right)}{2 e^{-( \rho_{c_1}+\rho_{c_2}) }\omega_{1} \omega_{2}}\right]+\frac{c}{12}  \log \left(1-2 \gamma ^2\right).
\end{align}
As explained earlier in \cref{ssec_pcft}, the entanglement entropy in the Janus deformed BTZ black hole geometry may be expressed in terms of undeformed entanglement entropy and a correction term as
\begin{align}\label{ee_gbtz}
	S_{\gamma}(A) & =\frac{c}{6}  \log \left[\frac{ \left(\omega_{1} \omega_{2}+\omega_{h}^2-\sqrt{\left(\omega_{1}^2-\omega_{h}^2\right) \left(\omega_{2}^2-\omega_{h}^2\right)} \cos r_{h}(t_2-t_1)\right)}{2 e^{-( \rho_{\epsilon_1}+\rho_{\epsilon_2}) } \omega_{1} \omega_{2}}\right]-\frac{c}{12}  \log \left(1-2 \gamma ^2\right) \notag \\
    & = \frac{c}{6}  \log \left[\frac{2 r_{\epsilon_1} r_{\epsilon_2} \left(\cosh r_{h}(x_{2}-x_{1})-\cosh r_{h}(t_{2}-t_{1})\right)}{r_{h}^2}\right]-\frac{c}{12}  \log \left(1-2 \gamma ^2\right),
\end{align}
where is the second step we utilize \cref{btz_trans} to transform back to the BTZ black hole coordinates. The results in \cite{Bak:2020enw} may be reproduced from \cref{ee_gbtz} by assuming $x_2=-x_1=L/2,~t_2=t_1$, and $r_{\epsilon_1},r_{\epsilon_2} \sim 1/2 \epsilon$, which gives us
\begin{align}
    S_{\gamma}(A) = \frac{c}{3} \log \left[ \frac{\sinh (r_h L/2)}{r_h \epsilon} \right] - \frac{c}{12}  \log \left(1-2 \gamma ^2\right)  .
\end{align}


\subsection{Holographic Entanglement Entropy} \label{ssec_btzbulk}

The BTZ black hole geometry in \cref{BTZ_metric} can also be constructed using the embedding space formalism. Similar to the previous case, the BTZ geometry may be embedded as a codimension-one hypersurface in $\mathbb{R}^{(2,2)}$ geometry as \cref{AdS Embedd Metric}. For the Janus black hole solution given by \cref{BTZ_metric}, the following embedding relations map the asymptotic regions of the deformed geometry to a codimension-one hyperboloid in $\mathbb{R}^{(2,2)}$
\begin{align}\label{BTZ_Embedding}
    T_1 = \sinh{\left(r_h t\right)} \sqrt{f(\rho ) \left(\frac{\omega_h^2}{\omega^2}-1\right)}, \qquad & \qquad T_2 = \frac{\omega_h}{\omega}\sqrt{f(\rho )},  \notag \\
    X_1 = \cosh{\left(r_h t\right)} \sqrt{f(\rho ) \left(\frac{\omega_h^2}{\omega^2}-1\right)},  \qquad & \qquad X_2 = \sqrt{f(\rho)-1}  . 
\end{align}
Using the embedding parametrizations given in \cref{BTZ_Embedding,AdS Length}, the length of the bulk geodesic connecting the boundary points $(t_1,\,\omega_1,\,\rho_{c_1})$ and $(t_2,\,\omega_2,\,\rho_{c_2})$ can be expressed as
\begin{align}
    \mathcal{L} = \log{\left[\frac{\sqrt{1-2\gamma^2} \left(\omega_1\omega_2 + \omega_h^2-\sqrt{\left(\omega_1^2-\omega_h^2\right) \left(\omega_h^2-\omega_2^2\right)} \cosh{(r_h (t_2-t_1))}\right)}{2e^{-(\rho_{c_1}+\rho_{c_2})}\; \omega_1 \omega_2}\right]},
\end{align}
and the holographic entanglement entropy may be computed using the HRT prescription as \cref{ee_btz}. Following the same arguments as in \cref{ssec_pbulk}, we can once again obtain the entanglement entropy in the Janus deformed BTZ black hole geometry using \cref{btz_trans} as 
\begin{align}
	S_{\gamma}(A) = \frac{1}{4 G_3}  \log \left[\frac{2 r_{\epsilon_1} r_{\epsilon_2} \left(\cosh r_{h}(x_{2}-x_{1})-\cosh r_{h}(t_{2}-t_{1})\right)}{r_{h}^2}\right]-\frac{1}{8G_3}  \log \left(1-2 \gamma ^2\right),
\end{align}
which matches exactly with the field theory results in \cref{ee_gbtz} on utilizing the Brown-Henneaux relation.


\section{Janus deformed AdS$_3$ black string geometry}\label{sec_bs}

In this section we consider the Janus deformation of the AdS$_3$ black string geometry dual to a CFT$_2$ defined on an AdS$_2$ black hole background. The undeformed bulk AdS$_3$ black string geometry is described by the metric \cite{Geng:2021mic,Geng:2022dua, Basu:2023jtf}
\begin{align}
	ds^2 =   d\rho^2+\cosh^2 \rho \left[ -\frac{\left(1 - \tfrac{u}{u_h}\right)}{u^2} dt^2 + \frac{du^2}{u^2(1 - \tfrac{u}{u_h})} \right] ,
\end{align}
where $\rho \in (-\rho_\infty, \rho_\infty)$. The asymptotic boundaries are located at $\rho= \pm \rho_\infty$, while $u_h$ describes the radius of the black string horizon. Each constant-$\rho$ slice of the bulk black string geometry corresponds to an eternal AdS$_2$ black hole which has two asymptotic boundaries. From the AdS$_3$/CFT$_2$ correspondence, the dual field theory is then a CFT$_2$ defined on an AdS$_2$ black hole background.

The Janus deformed AdS$_3$ black string geometry may be described by promoting the warping factor of each AdS$_2$ foliation to be dependent on the deformation parameter $\gamma$. Choosing the warping factor to once again be $f(\rho)$ described in \cref{f(rho)_def}, the deformed metric is given as
\begin{align}\label{black string}
	ds^2 =   d\rho^2+f(\rho) \left[ -\frac{\left(1 - \tfrac{u}{u_h}\right)}{u^2} dt^2 + \frac{du^2}{u^2(1 - \tfrac{u}{u_h})} \right]  ,
\end{align}
where $\rho \in (-\rho_c, \rho_c)$ and $\rho= -\rho_c \cup \rho_c$ is the asymptotic boundary. 

In what follows, we begin with the field theory computation of the entanglement entropy for a boosted single subsystem defined on the asymptotic boundary of a Janus deformed AdS$_3$ black string geometry. Subsequently, we will discuss the computations of the bulk holographic entanglement entropy.


\subsection{Entanglement Entropy} \label{ssec_bscft}

In this subsection, we compute the entanglement entropy for a boosted single subsystem $A=[(u_1,t_1),(u_2,t_2)]$ at the asymptotic boundary in the Janus deformed AdS$_3$ black string geometry utilizing the replica technique. Using \cref{black string} the metric at the asymptotic boundary is given as
\begin{align}
	ds^2 = f(\rho_c) \left[ -\frac{\left(1 - \tfrac{u}{u_h}\right)}{u^2} dt^2 + \frac{du^2}{u^2(1 - \tfrac{u}{u_h})} \right]  ,
\end{align}
where $f(\rho_c)= \sqrt{1-2 \gamma ^2} \left(\frac{e^{\rho_c}}{2 y}\right)^2.$ The above metric corresponds to an AdS$_2$ black hole and we have two copies of such geometry corresponding to the two asymptotic boundaries of \cref{black string}. The CFT$_2$ is then located in the AdS$_2$ black hole background, which is not conformally flat. As argued in the previous section, determining the twist field correlator in this curved background is not straightforward. Hence is necessary to map this field theory on the curved geometry to that on a flat background through an appropriate sequence of conformal transformations
\begin{align}\label{bs_tr1}
	u= u_{h}-\frac{u_h^3}{\omega  \bar{\omega}}, \qquad t= u_h \log \frac{\bar{\omega}}{\omega }.
\end{align}
The metric in these new coordinates becomes conformally flat,
\begin{align}
	ds^2=\frac{4 u_{h}^2 f(\rho ) }{\left(u_{h}^2-\omega  \bar{\omega}\right)^2}~d\omega  d\bar{\omega}.
\end{align}
It is possible to further conformally map this geometry to the upper-half-plane (UHP) through the transformations
\begin{align}\label{bs_tr2}
	\omega = \frac{u_{h}}{v -\frac{i}{2}}-i u_{h}, \qquad \bar{\omega}= \frac{u_{h}}{\bar{v}+\frac{i}{2}}+i u_{h},
\end{align}
where the metric now has the form
\begin{align}\label{UHP-metric}
	ds^2= \Omega^2 (v,\bar{v})~dv  d\bar{v}, \qquad \Omega (v,\bar{v})=\frac{\left(1-2 \gamma ^2\right)^\frac{1}{4} }{v -\bar{v}} e^{\rho }.
\end{align}
The entanglement entropy for a subsystem $A$ at the asymptotic boundary may be computed by utilizing the following two point twist field correlator
\begin{align}
	S(A)= \lim_{n\to 1} \frac{1}{1-n} \log  \langle \phi_n (u_1,t_1) \phi_n (u_2,t_2) \rangle.
\end{align}
Utilizing the form of the metric and the conformal factor described in \cref{UHP-metric}, the above expression may be written as
\begin{align}
	S(A)= \lim_{n\to 1} \frac{1}{1-n} \log \Omega(v,\bar{v})^{-\Delta_n} \langle \phi_n (v_1,\bar{v}_1) \phi_n (v_2,\bar{v}_2) \rangle_{\text{Flat}}.
\end{align}
Using the form of the two point function and substituting the conformal factor and taking the replica limit, we can obtain
\begin{align}
	S(A)=\frac{c}{6}  \log \left[\frac{(v_1-\bar{v}_{2})^2 e^{\rho_{c_{1}}+\rho_{c_{2}}}}{(v_1-\bar{v}_{1}) (v_2-\bar{v}_{2})}\right]+\frac{c}{12}  \log \left(1-2 \gamma ^2\right).
\end{align}
As explained in \cref{ssec_pcft}, the above expression of the entanglement entropy may be written in terms of the undeformed entanglement entropy with a correction term as 
\begin{align}
	S_{\gamma}(A)=\frac{c}{6}  \log \left[\frac{(v_1-\bar{v}_{2})^2 e^{\rho_{\epsilon_{1}}+\rho_{\epsilon_{2}}}}{(v_1-\bar{v}_{1}) (v_2-\bar{v}_{2})}\right]-\frac{c}{12}  \log \left(1-2 \gamma ^2\right).
\end{align}
By relating the UV–IR cutoffs in the CFT and the bulk, and subsequently transforming the above expression back into black string coordinates using \cref{bs_tr1,bs_tr2}, the entanglement entropy for a subsystem $A$ may be obtained as
 \begin{align}\label{ee_gbs}
 S_{\gamma}(A) & = \frac{c}{6} \log\!\left[\frac{ e^{\rho_{\epsilon_1}+\rho_{\epsilon_1}} \left(\Delta _1 \Delta _2-2u_h \sqrt{\Delta _1\,\Delta _2}  \cosh \left(\frac{t_1-t_2}{2 u_h}\right)+u_h^2\right)}{\left(u_h-\Delta _1\right) \left(u_h-\Delta _2\right)}  \right] - \frac{c}{12}\log{(1-2\gamma^2)}  ,
 \end{align}
where $\Delta_{i}= u_{h}-u_{i}$. Now, considering $t_1=t_2$, $u_1=u_2$ and $\rho_{\epsilon_1},\rho_{\epsilon_2} \sim 2/\epsilon$, we may obtain the entanglement entropy of a purely spatial subsystem described symmetrically about the interface as
\begin{align}\label{HEE_BS_sym}
 S_{\gamma}(A) & = \frac{c}{3} \log\!\left(\frac{2}{\epsilon}\right) - \frac{c}{12}\log{(1-2\gamma^2)} .
 \end{align}


\subsection{Holographic Entanglement Entropy} \label{ssec_bsbulk}

We now compute the holographic entanglement entropy in a Janus deformed AdS$_3$ black string geometry, where the metric is described by \cref{black string}. Similar to the previous cases, the Janus deformed AdS$_3$ black string geometry is asymptotically AdS in nature, and the patch in the vicinity of the asymptotic boundary (once again corresponding to large values of the hyperbolic angle $\rho$) may be embedded as a codimension-one hyperboloid in a $\mathbb{R}^{(2,2)}$ flat spacetime using the embedding relations 
\begin{align}\label{BS_Embed}
     T_1 = 2\sqrt{f(\rho)\left(\frac{u_h^2-u u_h}{u^2}\right)}\;\sinh{\left(\frac{t}{2 u_h}\right)},\qquad & \qquad T_2 = \sqrt{f(\rho )}\;\frac{ (2 u_h-u)}{u}, \notag \\ 
     X_1 = 2\sqrt{f(\rho)\left(\frac{u_h^2-u u_h}{u^2}\right)}\cosh{\left(\frac{t}{2 u_h}\right)}, \qquad & \qquad X_2 = \sqrt{f(\rho )-1} .
\end{align}
Consequently, using \cref{AdS Embedd Metric} we can reproduce the Janus deformed AdS$_3$ black string geometry in the large $\rho$ approximation.

We can now determine the geodesic length between two arbitrary boundary points $(u_1,\,t_1,\,\rho_{c_1})$ and $(u_2,\,t_2,\,\rho_{c_2})$ using \cref{AdS Length} and the embedding relations in \cref{BS_Embed} as
\begin{align}\label{length_blackstring}
    \mathcal{L} = \log \left[\frac{\sqrt{1-2 \gamma ^2} e^{\rho_{c_1}+\rho_{c_1}} \left(\Delta _1 \Delta _2-2u_h \sqrt{\Delta _1\,\Delta _2}  \cosh \left(\frac{t_1-t_2}{2 u_h}\right)+u_h^2\right)}{\left(u_h-\Delta _1\right) \left(u_h-\Delta _2\right)} \right] .
\end{align}
The holographic entanglement entropy may now be computed using the HRT prescription, and once again following the arguments in \cref{ssec_pbulk}, we may determine the holographic entanglement entropy in the Janus deformed black string geometry in terms of the undeformed coordinates and a correction term as 
\begin{align}\label{S_BS_deformed}
 S_{\gamma}(A) & = \frac{1}{4G_3} \log\!\left[\frac{ e^{\rho_{\epsilon_1}+\rho_{\epsilon_1}} \left(\Delta _1 \Delta _2-2u_h \sqrt{\Delta _1\,\Delta _2}  \cosh \left(\frac{t_1-t_2}{2 u_h}\right)+u_h^2\right)}{\left(u_h-\Delta _1\right) \left(u_h-\Delta _2\right)}  \right] - \frac{1}{8G_3}\log{(1-2\gamma^2)}  .
 \end{align}
Applying the Brown-Henneaux relation to the above equation gives us \cref{ee_gbs}.


\section{Summary and Discussions}\label{sec_summary}

In this article, we investigate the holographic entanglement entropy in a family of Janus deformed AdS$_3$ geometries dual to ICFT$_2$s. To this end, we begin with the Janus deformed Poincaré AdS$_3$ patch and consider a boosted single interval asymmetric about the interface. We compute the entanglement entropy and the Janus induced correction from both field theoretic and bulk perspectives, and find an agreement between the results. We further extend the analysis to Janus deformed BTZ black hole AdS$_3$ black string geometries, where we once again find an exact match between the field theory and bulk results.

For the field theory computation, we describe a conformal mapping approach involving the effective boundary metric induced by the bulk deformation, as an alternative to the conformal perturbation techniques described in the literature. This method treats the boundary theory as a two-dimensional CFT on a conformally flat background. The entanglement entropy is then evaluated using the replica technique, incorporating the correction due to the Janus deformation through the Weyl conformal factor. For the corresponding bulk computations we utilize an embedding space formalism to obtain the length of the RT/HRT surface. Through our results we demonstrate that the Janus induced correction to the entanglement entropy assumes a universal form dependent only on the deformation parameter $\gamma$, independent of the subsystem size and location.

Our analysis inspires several open questions for future investigations. It would be interesting to understand if our field theory and bulk computational methods may be utilized to examine dynamical or time-like Janus solutions \cite{Bak:2007, Nakaguchi:2015, Suzuki:2025}, potentially shedding light on holographic quenches and causal propagation of entanglement in such geometries. Another exciting direction would be to investigate the correction to mixed state entanglement and correlation measures due to Janus deformation, a problem yet unexplored in the literature. Furthermore, applying our framework to investigate quantum information aspects of the black hole information paradox, holographic complexity, and traversable wormholes \cite{Bak:2021, Auzzi:2021, Auzzi:2022, Kawamoto:2025oko} would be a natural extension of our work. We leave such interesting issues for future considerations.


\section*{Acknowledgments}
We are thankful to A. Karch for useful discussions and clarifications. Ankit Anand is financially supported by the Institute's postdoctoral fellowship at Indian Institute of Technology Kanpur. Himanshu Chourasiya is supported by the Fellowship for Academic and Research Excellence (FARE) at IIT Kanpur.

\bibliographystyle{utphys.bst}
\bibliography{ref}
\end{document}